\documentclass{article}

\usepackage{arxiv}

\usepackage[utf8]{inputenc} 
\usepackage[T1]{fontenc}    
\usepackage{hyperref}       
\usepackage{url}            
\usepackage{booktabs}       
\usepackage{amsfonts}       
\usepackage{nicefrac}       
\usepackage{microtype}      
\usepackage{lipsum}
\usepackage{graphicx}
\graphicspath{ {./images/} }

\title{What information should be shared with whom “before and during training”?}

\author{
 Haydn Belfield \\
  Leverhulme Centre for the Future of Intelligence, University of Cambridge\\
  \texttt{hb492@cam.ac.uk} \\
}

\begin{document}
\maketitle
\begin{abstract}
In the Frontier AI Safety Commitments, sixteen companies committed to “Assess the risks posed by their frontier models or systems across the AI lifecycle, including […] as appropriate, before and during training” (I) and to “Provide public transparency on the implementation of the above (I-VI), except insofar as doing so would increase risk or divulge sensitive commercial information to a degree disproportionate to the societal benefit. They should still share more detailed information which cannot be shared publicly with trusted actors, including their respective home governments or appointed body, as appropriate” (VII). This short paper considers what information should be shared with whom before training begins. What information should be shared publicly and what only with trusted actors such as home governments? Sharing such information before a frontier training run can build shared awareness and preparedness, can improve risk assessment and management, and can contribute to greater predictability and accountability. 

Companies could share certain information publicly or with trusted partners before the frontier training run including:
\begin{itemize}
   \item Expected dates of beginning and end of training;
\item Expected compute used (in FLOP);
\item Description of the pre-training dataset(s), for example to what extent they include biological sequence data and/or steps taken to prevent data poisoning;
\item Expected capability level of the frontier model or system, including an assessment of potential risks (such as ‘AI Safety Levels’ or ‘Critical Capability Levels’), and whether this capability will approach any risk threshold; 
\item How the company will monitor progress, capabilities and risks during training so that they are able to “further develop” the system and model “only if they assess that residual risks would stay below the thresholds” (IV);
\item Location of the large-scale computing cluster(s) used for the frontier training run, ownership of that cluster(s), primary energy source (nuclear, gas, etc.) and amount of total computing power available in each cluster;
\item Physical, personnel and cybersecurity steps taken; and
\item Which internal groups have been tasked – and which external bodies have been contracted – to carry out evaluations and red-teaming, what level of resources and support they have available to do so, and how many weeks and months those bodies will have to do so.
 \end{itemize}
\end{abstract}


\section{Introduction}
At the AI Seoul Summit 2024, sixteen companies from across the world voluntarily agreed to the Frontier AI Safety Commitments. This paper examines the intersection between two of these Commitments:
\begin{itemize}
   \item[]“Assess the risks posed by their frontier models or systems across the AI lifecycle, including […] as appropriate, before and during training” (Commitment I) 
   \item[]“Provide public transparency on the implementation of the above (I-VI), except insofar as doing so would increase risk or divulge sensitive commercial information to a degree disproportionate to the societal benefit. They should still share more detailed information which cannot be shared publicly with trusted actors, including their respective home governments or appointed body, as appropriate” (Commitment VII). 
   \end{itemize}

There are existing requirements to share some information before training. For example, there are requirements to share information in the Executive Order 14110 and related Reporting Requirements \cite{bureau_of_industry_and_security_establishment_2024}
and the AI National Security Memorandum \cite{national_security_council_memorandum_2024}. Companies must provide similar information under Annexes IV and XI of the EU AI Act and the draft Code of Practice. Similar requirements may be included in the forthcoming UK frontier AI regulation Bill. Some elements of information sharing before training have been briefly discussed before in the literature on frontier AI \cite{han_voluntary_2022,anderljung_frontier_2023,dsit_emerging_2023,koessler_risk_2024,robinson_transforming_2024,metr_common_2024,irving_safety_2024, microsoft_global_2024}.

This paper examines some overall benefits of information sharing before training, then three challenges and mitigations, and finally what specific information could be shared with whom – weighing these benefits, challenges and mitigations. This is a short paper that is intended to enable discussion and iteration.

\section{Benefits of information sharing before training}
\label{sec:headings}

Sharing information before training can build shared awareness and preparedness, can improve risk assessment and management, and can contribute to greater predictability and accountability. 

Information sharing before training can build shared awareness and preparedness. Even a year ago, states and the public had no warning that a company was training a powerful new model – the first that states would hear about it would be the public announcement of the corporate blog post. Over time, this could have led to a situation where governments were playing ‘catch up’, scrambling to respond and only responding in a reactive way. It could even have led to overreaction and miscalibrated clampdowns. Sharing information before training can build shared awareness and give governments and the public more warning, avoiding these unhelpful outcomes. Moreover, this can help with preparedness. Evaluators and red-teamers – companies, NGOs and the AI Safety Institutes – can better prepare to carry out evaluations and red-teaming if they know when training will be completed. Governments may also want to take other steps to prepare for release of a frontier model – such as preparing cybersecurity responses, police responses (if capabilities of the model are misused criminally), preparing public services (such as schools or universities), and/or employment and skills retraining services. This is all easier to prepare for with a pre-training ‘heads-up’.

Information sharing before training can improve risk assessment and management. Information on the expected capability level of the trained model and whether it is expected to cross any risk thresholds is an important part of risk assessment. This could be cross-referenced with information on expectations of duration of training run, compute use and dataset size - given scaling laws these should have some proportion to one another, and can inform expected capability and risk levels \cite{sastry_computing_2024}. Description of datasets can inform risk assessment too. For example, if the pre-training dataset(s) contains biological sequence data this could inform the assessment of whether the trained model may have dual-use biological capabilities. It can also help different actors manage risks before and during the training period – such as helping national cyber agencies deter and prevent cyber espionage or sabotage of the datasets or cluster.

Information sharing before training can contribute to greater predictability and accountability. Companies stating clearly what capability level and risk thresholds they predict, and providing further information that can be cross-referenced with that, can help governments and society at large better predict the capabilities of new frontier models and systems and the impacts they might have on economies and societies. Accountability can be furthered by companies stating clearly before training which internal and external teams will be carrying out evaluations and red-teaming and what support they will have. This can be cross-referenced after training in the evaluation stage, and then again at publication/release: did the company follow through with what they said they would do? Companies have committed publicly to develop their systems responsibly, safely and ethically. Assuring the public that their systems have been tested by independent third parties who have been given the time and resources to do their job properly is a way for these companies to demonstrate accountability and earn trust \cite{brundage_toward_2020,avin_filling_2021}. Information sharing could establish best practice and help all companies meet it. For example, a company that shared information that was accurate, clear and precise could be compared to one that shared information that was limited, confusing or vague.

Finally and more generally, governments want reassurance that no unsafe, insecure, dangerous or secret frontier training runs are being conducted in their jurisdictions. They want to know whether, where, when and how these frontier training runs are being conducted and whether their legitimate safety and security concerns are being properly assessed and managed. Sharing information before training with the public and/or trusted actors such as home governments can help provide that reassurance.

\section{Challenges and mitigations}

Sharing information before training could pose three challenges: it could be burdensome or – as the Frontier Safety Commitments acknowledge – it could “increase risk or divulge sensitive commercial information to a degree disproportionate to the societal benefit”. These should be taken seriously. Luckily, mitigations are possible.

One concern could be that sharing information could be burdensome. However, a frontier training run is a significant, expensive and lengthy undertaking for a company preceded by a lot of planning, internal checks, documentation and risk assessment. Transparency about some aspects of that process should not represent a significant additional burden. The cost and time of generating documentation would, for example, be a small fraction of the cost and duration of a frontier training run \cite{laurer_clarifying_2021,the_future_society_eu_2023}. The main reason for this is that companies will already hold all of the information discussed in this paper: for example, they know when, where and how they will conduct their frontier training run, and to fulfil their other Frontier Safety Commitments they should have predictions of capability levels and risk thresholds, and plans for evaluation and red-teaming.

Serious attention should be paid to when sharing information publicly might “increase risk or divulge sensitive commercial information to a degree disproportionate to the societal benefit”, and so should only be shared with trusted actors such as home governments.

On some topics where sharing information could “increase risk”, limited information can be shared publicly while full information should only be shared with trusted actors such as home governments. For example, a company could publicly state whether it deems the cybersecurity of the data centre and pre-training team to be sufficient to fulfil their Commitments, while sharing further information with trusted actors such as home governments on e.g. what RAND Security Level (L1-L5) \cite{nevo_securing_2024} they are operating at as well as further information on physical, personnel and cybersecurity steps taken to protect the training process and model weights.

It will be important to consider feedback from companies, researchers and civil society on what information might “divulge sensitive commercial information to a degree disproportionate to the societal benefit”. For example, in competition policy cost structures and release dates are sometimes viewed as sensitive commercial information. If sensitive commercial information is shared publicly or with governments (rather than privately with a select group of competitors), then this kind of information-sharing should not raise any antitrust or competition law concerns \cite{hua_ai_2021}. This feedback should inform what information is shared publicly and which should only be shared with trusted actors such as home governments.  It is important to note that the Commitment calls for proportionality - weighing up how commercially sensitive the information is and how much societal benefit would derive from public transparency about this information. These costs and benefits must be weighed against each other.

\section{Specific information that could be shared before training}

This section provides more detail on specific forms of information that could be shared before training – weighing benefits, challenges and mitigations. Companies could share certain information publicly or with trusted partners before the frontier training run, including:
\begin{itemize}
   \item Expected dates of beginning and end of training;
\item Expected compute used (in FLOP);
\item Description of the pre-training dataset(s), for example to what extent they include biological sequence data and/or steps taken to prevent data poisoning;
\item Expected capability level of the frontier model or system, including an assessment of potential risks (such as ‘AI Safety Levels’ or ‘Critical Capability Levels’), and whether this capability will approach any risk threshold; 
\item How the company will monitor progress, capabilities and risks during training so that they are able to “further develop” the system and model “only if they assess that residual risks would stay below the thresholds” (IV); and
\item Location of the large-scale computing cluster(s) used for the frontier training run, ownership of that cluster(s), primary energy source (nuclear, gas, etc) and amount of total computing power available in each cluster;
\item Physical, personnel and cybersecurity steps taken;
\item Which internal groups have been tasked – and which external bodies have been contracted – to carry out evaluations and red-teaming, what level of resources and support they have available to do so, and how many weeks and months those bodies will have to do so.
 \end{itemize}

\textbf{Expected dates of beginning and end of training.}\par

This could help governments by building shared awareness of when a frontier training run will be conducted, of the speed and nature of progress, and of the capabilities and risks of frontier AI models and systems. It could help government cybersecurity professionals better support companies. It could help AI Safety Institutes plan their work, and ensure they have capacity available for evaluations and red-teaming when training is finished, even potentially speeding up the evaluation process. Duration also provides evidence that can be cross-referenced with other sources of information to estimate capability levels and risk thresholds.

Companies may view this information as commercially sensitive and be wary of sharing it publicly – as it could give their competitors insight on release dates, capabilities/performance and cost structures. Companies could be concerned that competitors could change their development and release plans to try and beat or compete with those dates. There would be substantial societal benefits to sharing this information publicly – it could enable a wide range of actors to prepare – from schools, and universities to job centres and retraining providers. These costs and benefits must be weighed against one another.\newline

\textbf{Expected compute used (in FLOP).}

This could help governments know when a frontier training run will cross a compute threshold. Frontier regulation often uses compute thresholds as a input-based proxy for additional scrutiny. Knowing whether such thresholds will be crossed and further scrutiny triggered would be important and helpful information for governments (and the wider public). Moreover, compute can be cross-referenced with other sources of information to estimate capability levels and risk thresholds.

Companies may view this information as commercially sensitive and be wary of sharing it publicly as it could give their competitors insight on capabilities/performance and cost structures, given that training compute is a significant aspect of overall cost. There would be substantial societal benefits to sharing this information publicly as it would allow better assessment of the predicted capability of the trained model, given scaling laws. It would also give insight into energy usage and likely emissions. These costs and benefits must be weighed against one another.\newline

\textbf{Description of the pre-training datasets, for example to what extent they include biological sequence data and/or steps taken to prevent data poisoning.}

This could help governments understand specific capabilities and risks of the model – for example to assess the risk of CBRN (chemical, biological, radiological, and nuclear) capability gains or whether the model may be compromised by cyber criminals. Moreover, size of datasets can be cross-referenced with duration and compute use to estimate capability levels and risk thresholds.

Companies have often claimed that their pre-training dataset(s) are commercially sensitive and have been wary of sharing information about them publicly. Access to high-quality datasets that their competitors do not have or do not even know about can give companies advantage. These companies may have invested in deals with rights-holders or may have invested in developing their own datasets. However, it has also been alleged that greater transparency could reveal scraped copyrighted material in the dataset(s) and provide evidence for lawsuits. There could be substantial societal benefits of providing information publicly to inform that debate, or to build shared awareness about likely capabilities and possible risks of a given model. These costs and benefits must be weighed against one another.\newline

\textbf{Expected capability level of the frontier model or system, including an assessment of potential risks (such as ‘AI Safety Levels’ or ‘Critical Capability Levels’), and whether this capability will approach any risk threshold.}

This could help governments better prepare for capability and risk increases. Companies have committed to assess capability levels and risk thresholds and sharing these assessments would be helpful. This could be cross-referenced with the previous three points: information about training duration, expected compute usage, and dataset size. 'Scaling law' hypotheses imply the optimal combination of these inputs. If all of these four line up that can give other actors confidence that information about each are consistent and accurate. By contrast, if these four do not line up this could prompt further questions.  

Companies may view this information as commercially sensitive and be wary of sharing it publicly – as it could give their competitors insight on capabilities/performance and cost structures. There would be substantial societal benefits to sharing this information publicly as it would allow better assessment of predicted capability and risks, promote predictability and accountability, and build and encourage best practice in assessment of capabilities and risks. These costs and benefits must be weighed against one another.\newline

\textbf{How the company will monitor progress, capabilities and risks during training so that they are able to “further develop” the system and model “only if they assess that residual risks would stay below the thresholds” (IV).}

This could help reassure governments (and the wider public) that companies are properly monitoring for unexpected surprises during training. Unexpected changes can happen during training: frontier training runs and scaling ‘laws’ are not an exact science. Even with the best efforts of companies, governments and experts to predict capabilities and risks associated with a given level of inputs (in terms of compute, dataset size, duration, etc.), perfect prediction is not possible and surprises can happen. Companies should therefore be monitoring progress, capabilities and risks during training to monitor for these kinds of surprises. Governments and the public will want reassurance that they are doing so properly. Transparency can provide that reassurance, as well as sharing best practice.

This information should be carefully calibrated to ensure that sharing it does not “increase risk or divulge sensitive commercial information to a degree disproportionate to the societal benefit”, for example by highlighting possible cyber vulnerabilities.\newline

\textbf{Location of the large-scale computing cluster(s) used for the frontier training run, ownership of that cluster(s), primary energy source (nuclear, gas, etc.) and amount of total computing power available in each cluster.}

This could help reassure governments that no frontier training runs are being conducted without their knowledge and help governments better assess (cyber)security threats to companies in their jurisdictions. Ownership information would provide clarity on which company is responsible and accountable for training, and is best placed to answer any further questions. Primary energy source information would help estimate emissions of the training run.

Companies may view this information as commercially sensitive and be wary of sharing it publicly – as it could give their competitors insight on cost structures. There would be substantial societal benefits to sharing this information publicly as it would allow better assessment of predicted capability of the trained model, given scaling laws. It would also give insight into likely emissions. If companies' power sources can be compared, this could encourage a race to the top on reducing emissions. These costs and benefits must be weighed against one another.

There may be a concern that sharing full information publicly might increase risk. For example, one could be concerned that revealing which specific cluster the frontier training run will be conducted on may increase risk of cyberattack. However, sizes of frontier clusters are growing substantially. A 1 GW cluster is really only economical for a frontier training run – it would be too big to be optimal for inference. As time goes on it will likely be increasingly obvious where a frontier training run is conducted. Moreover, energy sources can often be observed from publicly available sources such as satellite data, e.g. onsite gas turbines or connections to a nearby power plant. Nevertheless, insofar as sharing full information publicly could increase risk, then full information should only be shared with trusted actors such as home governments. \newline

\textbf{Physical, personnel and cybersecurity steps taken.}

This could help reassure governments that AI companies are taking the necessary steps to protect the training process and model weights from sabotage, espionage and theft. It could also help governments understand where companies may need more support. For example, getting to RAND Security Level 5 (the highest level) is technically and organisationally difficult and could be quite a change to existing ways of working at AI companies: more restricted access to hardware, more vetting of personnel, and more restricted access to software. Companies may need additional support in these areas. As the BIS Establishment of Reporting Requirements notes:

\begin{itemize}
    \item[]“the U.S. Government must minimize the vulnerability of dual-use foundation models to cyberattacks. Dual-use foundation models can potentially be disabled or manipulated by hostile actors, and it will be difficult for the U.S. Government to rely on a particular model unless it can determine that the model is robust against such attacks. Accordingly, the U.S. Government needs information about the cybersecurity measures that companies developing dual-use foundation models use to protect those models, as well as information about those companies’ cybersecurity resources and practices.” (Bureau of Industry and Security, 2024, p. 6)
\end{itemize}

Sharing information publicly on high standards for security could build best practice and help other companies meet it. If commercially sensitive information is shared directly and privately between competitors this may raise competition concerns – but these can be mitigated by being limited to only what is necessary, or sharing this information through a third-party or through a government process.  

Sharing full information publicly could increase risk by revealing vulnerabilities to attackers. Publicly sharing information should therefore be restricted to only what is necessary to affirm that the security is sufficient to fulfil a company’s Frontier Safety Commitments. In their Frontier Safety Frameworks companies have occasionally given information on what RAND Security Level they are operating at – this could continue if it does not increase risk.\newline

\textbf{Which internal groups have been tasked – and which external bodies have been contracted – to carry out evaluations and red-teaming, what level of resources and support they have available to do so, and how many weeks and months those bodies will have to do so.}

This could help reassure governments and the public that AI companies are fulfilling their promises to develop AI responsibly, and that systems have been rigorously tested by independent third parties. For example, if a company stated that they were contracting no external bodies, or providing them with inadequate access, resources, support and time to carry out evaluations and red-teaming, then society and governments might have reasonable questions about how rigorous the process was. It would also help AI Safety Institutes, evaluators and red-teamers better prepare – as they would be notified or contracted before training rather than later in the process. It would also establish and build best practice and encourage a race to the top - as companies experiment, find and share the best way to run these evaluations. For example, one company could commit to run uplift studies on their new model. This is akin to randomised control trials: it produces better quality evidence than automated evaluations, but can cost more and take longer. Other companies could see that commitment and seek to match or better it.

It seems unlikely that companies will view this information as commercially sensitive or be wary of sharing it publicly before training, as after training they include this information in papers or System Cards. There may be a concern that it could identify external groups as targets for cyberattacks, increasing risk. However, there are currently only a few AI Safety Institutes and evaluators/red-teamers in the world, and specifying which have been contracted would likely have marginal or no impact on their level of risk. Moreover, model access has been structured to minimise security risks.

\section{Conclusion}

This short paper has discussed the benefits and challenges of sharing information before and during the training run with the public and trusted actors such as home governments. Sharing this information should not represent an additional burden as it is already in the hands of companies. Serious attention should be paid to when and to what extent sharing full information publicly might “increase risk or divulge sensitive commercial information to a degree disproportionate to the societal benefit”, and so should only be shared with trusted actors such as home governments.

This information could include expected dates of beginning and end of training, expected compute and a description of the dataset(s), all of which can be cross-referenced with the expected capability levels and risk thresholds – and how capabilities and risks will be monitored during training. Information could also include the location, ownership, energy source and compute capacity of the cluster(s) and physical, personnel and cybersecurity steps taken to protect it. Finally, it could include which internal and external bodies will be carrying out evaluations and red-teaming.

Sharing some kinds of information before training can help companies, governments, evaluators and red-teamers, and wider societies be more aware of and prepare for progress, capabilities and, risks and changes; can help improve risk assessment and management; and can help to advance greater predictability and accountability. 

\section*{Acknowledgments}\label{sec:acknowledgments}

Thanks to participants and organisers of the 2024 Conference on Frontier AI Safety Commitments for feedback on earlier versions of this paper.

\bibliographystyle{unsrt}  
\bibliography{what_info_references}  






\end{document}